\documentclass[twocolumn,epsfig,showpacs,floatfix]{revtex4}
\usepackage{epsfig,colordvi}

\usepackage{epsfig}
\usepackage{graphicx}
\usepackage{graphics}

 \def\be{\begin{equation}}
 \def\ee{\end{equation}}
 \def\bea{\begin{eqnarray}}
 \def\eea{\end{eqnarray}}
 \def\bean{\begin{eqnarray*}}
 \def\eean{\end{eqnarray*}}

\begin{document}
\title{Centrality Dependence of Strangeness Enhancement in
Ultrarelativistic Heavy Ion Collisions - a Core-Corona Effect}
\author{J. Aichelin, K. Werner}
\affiliation{SUBATECH, Laboratoire de Physique Subatomique et des
Technologies Associ\'ees, \\
Universit\'e de Nantes - IN2P3/CNRS - Ecole des Mines de Nantes \\
4 rue Alfred Kastler, F-44072 Nantes, Cedex 03, France\\}
\begin{abstract}
In ultrarelativistic heavy ion collisions, the multiplicity of
multi-strange baryons per participating nucleon increases with
centrality in a different fashion for different systems and
energies. At RHIC, for copper+copper (CuCu) collisions the increase
is much steeper than for gold-gold (AuAu) collisions. We show that
this system size dependence is due to a core-corona effect: the
relative importance of the corona as compared to the core
(thermalized matter) contribution varies and the contribution of a
corona nucleon to the multiplicity differs from that of a core
nucleon. $\phi$ mesons follow - as all hadrons - the same trend, but
the difference between core and corona multiplicity is relatively
small, and therefore the CuCu and AuAu results are quite similar.
This simple geometrical explanation makes also a strong case in
favor of the validity of Glauber geometry in the peripheral regions
of ultrarelativistic heavy ion collisions, which is crucial for
understanding the early evolution of the system.
\end{abstract}
\pacs{}
\date{\today} \maketitle
Even before the first relativistic heavy ion beam has been delivered,
the enhancement of the production of strange particles has been
considered as a possible signal for the existence of a plasma
composed of quarks and gluons (QGP) \cite{koch}. This enhancement
may occur if due to compression the chemical potential of the up
and down quarks becomes that large that the system creates
preferably strange quarks and antiquarks which materialize finally
into strange hadrons. Since then strangeness enhancement is one of
the hot topics in the analysis ultrarelativistic heavy ion
collisions.

In the meantime heavy ion experiments have revealed that the multiplicity
per participating nucleon of (multi)strange baryons is up to 20
times larger than that observed in pp collisions per participating
proton at the same energy and that this enhancement is strongly
centrality dependent \cite{tak}. The observed multiplicities in
central collisions of heavy systems at RHIC and SPS energies can be
well described assuming that the hadrons are in statistical
equilibrium at a temperature close to the critical temperature
predicted by lattice gauge calculations
\cite{beca1,Andronic:2002pj,BraunMunzinger:2003zd,Andronic:2005yp}.
The chemical potential which is obtained by a fit to the data is
small and therefore the relation to the originally predicted
enhancement \cite{koch} is not evident.

Despite of many efforts, the centrality dependence of this
experimentally observed enhancement has not yet found a generally
accepted explanation. In (grand) canonical statistical models the
enhancement for symmetric systems is not dependent on the
centrality. It has been advocated that the increase with centrality
may be due to finite size effects (canonical suppression), but this
gives only a sizable effect for very small volumes
\cite{Redlich:2001kb}. To agree with data one needs to assume that
the effective volume depends on the number of participants as
$N_{\rm part}^{1/3}$, which is  in contradiction to the observation
at lower beam energies and which has not found a physical
explanation yet. It has also been shown that the centrality
dependence can by parameterized as a sum of two terms, one
proportional to the number of participants, $N_{part}$, the other
proportional to the number of binary collision, $N_{bin}$,
\cite{cai}, however no physical interpretation of this dependence
has been given.

Picking up an older idea \cite{ka} recently it has been proposed
that the dependence of the multiplicity on the centrality has a
geometrical origin \cite{klaus}: In this model one distinguishes
between a dense volume area, referred to as {\it core}, and a low
density peripheral region, referred to as {\it corona}. Only the
core participates in a collective expansion and produces particles
in a statistical manner, whereas the corona region consists mainly
of hadrons produced in nucleon-nucleon collisions. Employing this
idea in the framework of one of the most advanced models for the
simulation of heavy ion reaction, EPOS 1.9, it has been shown that
the difference in the centrality dependence between AuAu at RHIC and
PbPb at the SPS can be explained in a natural way with this
core-corona hypothesis. Later the authors of ref. \cite{beca1}
advanced a similar idea.

The parameters for this model have been determined to describe the
AuAu data. Therefore, in order to validate this approach, data for
other systems are necessary. The recent analysis of the RHIC CuCu
data at the same beam energy provides such a data set: in the
core-corona approach the particle multiplicities for different
systems are determined by the relative size of the core and the
corona. This is given by geometry. Therefore, the centrality
dependence of the particle multiplicities for CuCu can be predicted
without any new parameter.

The purpose of this article is twofold: a) to demonstrate that these
CuCu data are reproduced by Epos 1.9 and b) to show that the physics of
the centrality dependence of the different (multi)strange hadrons in
different systems can be understood in a simple model based on the
geometrically determined relative size of corona and core.

Without any calculation some basic features of the CuCu data find
their natural explanation in this core-corona approach. The observed
identical enhancement (as compared to pp collisions) for central
CuCu and AuAu collisions is due to the fact that central collisions
are to almost 100\% core (and the cores in CuCu and AuAu behave
identically at the same beam energy). Consequently, the slope of the
enhancement as a function of $N_{part}$ is larger in CuCu. With
decreasing centrality the corona contributes more and more, up to
$N_{\rm part}=2$ which is pure corona ($ = pp$). In pp collisions
the suppression (as compared to high energy $e^+e^-$ strings) of
strange baryons increases with the number of strange quarks which it
contains \cite{dresch}.  Therefore the production of multistrange
baryons is much more important in the core (thermal production) as
compared to the corona (string decay). Dividing the central yield in
heavy ion reactions by the yield in pp collisions the enhancement
factor increases with the number of strange quarks contained in the
baryon. This experimental observation has been baptized strangeness
enhancement although, physically speaking, it is a due to
strangeness suppression in pp.

This picture also explains naturally why strangeness enhancement is
more pronounced at SPS as compared to RHIC.  There are two reasons:
First, at lower energies the corona contribution relative to the
core contribution is larger over the whole range of $N_{\rm part}$.
Second, as has already been discussed quite a while ago
\cite{dresch}, the average string mass in pp collisions is rather
small and even at $\sqrt{s}= 200 \ {\rm AGeV}$ more than 30\% of the
strings have an energy smaller than the threshold for $\Omega$
production. Therefore the multi(strange) baryon/pion ratio is
suppressed as compared to a very energetic $e^+e^-$ string, and this
suppression increases with decreasing proton energy.

In order to understand the physics of the centrality dependence of
the multiplicities we develop first a toy model, which reproduces
nevertheless {\it quantitatively} the experimental results. Later will we
employ the sophisticated three-dimensional implementation of the
core-corona concept in the EPOS model, which confirms the conclusion
of this simple model.

In our toy model, we separate the wounded (or participating)
nucleons, $N_{part}$, into two classes. Those which  have only
scattered once are considered as "corona participants", which
hadronize like a string in pp collisions. The others form the
fireball and the particle multiplicity is given by phase space.
Defining $f(N_{\rm core})$ to be the ratio of participating "core
nucleons" to all participants, and assuming that the multiplicity
per wounded core nucleon, $M_{\rm core}$, and the one per wounded
corona nucleon, $M_{\rm corona}$, does neither depend on $N_{part}$
nor on the system size, the centrality dependence of hadron
multiplicities is given as
\begin{eqnarray}
 & &M^i(N_{\rm part})\label{eq1}\\
 & & = N_{\rm part}\,\big[f(N_{\rm core})\cdot M^i_{\rm core}+(1-f(N_{\rm core}))\cdot M^i_{\rm corona}\big]
\,\nonumber\end{eqnarray}
where $i$ refers to the hadron species. It is important to notice
that for a given total number of wounded nucleons, $f$ depends  on
the size of the interacting system. For RHIC experiments at
$\sqrt{s}$ = 200 and 62 AGeV, $f(N_{\rm core})$  has recently been
published \cite{Timmins:2008jm}. However, there the calculations are
done for the center of the centrality bins, whereas here we use a
Glauber model calculation which averages $f(N_{\rm core})$ over the
centrality bins. This gives different results in more peripheral
reactions. $M_{\rm core}$ can either by taken from statistical model
fits or can be directly determined from experiment. For our
calculation, we fix $M_{\rm core}$ by applying eq. \ref{eq1} to the
most central Au+Au or Pb+Pb data point, by replacing the left hand
side of eq. \ref{eq1} by the experimental value. $M_{\rm corona}$ is
given as half of the multiplicity measured in pp collisions. Once these
parameters are fixed, the centrality dependence of $M^i$ is
determined by eq. \ref{eq1}. Especially the centrality dependence of
the lighter Cu+Cu system follows then without any further input.

\begin{table}[hbt]
\centering
\begin{minipage}[c]{0.5\textwidth}
\begin{tabular}{|c|c|c|c|c|}
\hline cm energy & &$M_{\rm corona}$ & $M_{\rm core}$ &$M_{\rm core}^{\rm stat}$\\
\hline
200& $\Lambda$ & 0.0193 $\pm$ 0.0018 & 0.05 & 0.043\\
&$\bar\Lambda$&0.018$\pm$ 0.0017&0.038&0.037\\
&$\Xi$&0.0013$\pm$ 0.0005&0.0067&0.0052\\
&$\bar\Xi$&0.00145$\pm$ 0.0005&0.0056&0.0046\\
&$\Omega+\bar\Omega$&0.00017$\pm$0.0001&0.0016&0.0016\\
&$\phi$&0.009$\pm$0.0015&0.024&0.019\\
\hline 62&$\Lambda$&0.0112&0.047& 0.04\\
&$\bar\Lambda$&0.00816&0.022&0.022\\
&$\Xi$&0.00088&0.0055&0.0044\\
&$\bar\Xi$&0.00074&0.0036&0.0030\\
&$\Omega+\bar\Omega$&0.0001&0.0012&0.00123\\
&$\phi$&0.0065&0.024&0.014\\
\hline
NA49&$\Lambda$&as &0.033&\\
17.2&$\bar \Lambda $&for&0.00514&\\
&$\Xi$&NA 57&0.005&\\  \hline
NA57 &$\Lambda$&0.0137$\pm$0.0002&0.057&0.043\\
17.2&$\bar\Lambda$&0.0044$\pm$ 0.00008&0.0073&0.0060\\
&$\Xi$&0.0006$\pm$ 0.00004 &0.00648&0.0036\\
&$\bar\Xi$&0.00027$\pm$ 0.00004&0.00158&0.0010\\
&$\Omega+\bar\Omega$&0.000064$\pm$ 0.000024&0.00125&0.00065\\
\hline
\end{tabular}\vspace*{1ex}
\end{minipage}
\caption{Corona, $M_{corona}$, and bulk, $M_{\rm core}$, multiplicity
per wounded nucleon as well as statistical model predictions
\cite{Andronic:2005yp}  for $M_{\rm core}$.} \label{tab}
\end{table}
The values of $M_{\rm core}$ and $M_{\rm corona}$ as well as the
thermal predictions \cite{Andronic:2005yp} of $M_{\rm core}^{\rm
stat}$ are summarized in table 1. All multiplicities in this paper
refer to $dn/dy|_{y=0}$.  To determine $M_{\rm corona}$ we use the
data of \cite{rhicpp} for $\sqrt{s}=200 \ {\rm AGeV}$ and EPOS 1.9
results for $\sqrt{s}=62 \ {\rm AGeV}$. $M_{corona}$ is determined
from pBe for the NA57 and the NA49 data. The experimental results
can be found in \cite{rhicpp,RHIC,na49,na57,na49b} and are partially
still preliminary. We include in our analysis the centrality
dependence of the $\phi$ meson because in the past it has been
argued that it reflects the increased production of s-quarks in a
dense medium \cite{chen}.

\begin{figure}[ht]
\begin{center}
\hspace*{-0.5cm}
\includegraphics[scale=.42,angle=-90]{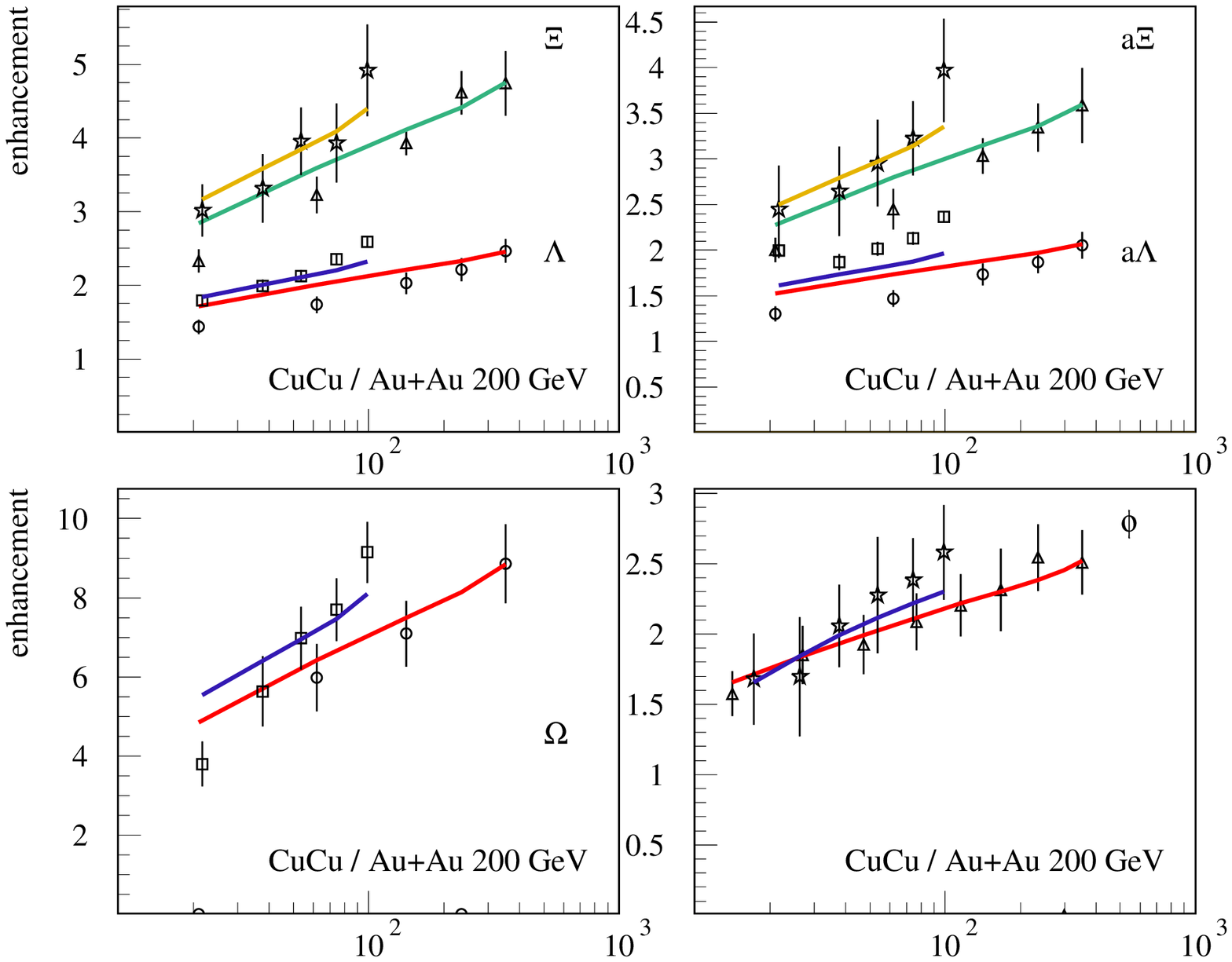}
\vspace*{-1.1cm}\par
\hspace*{-0.5cm}
\includegraphics[scale=.42,angle=-90]{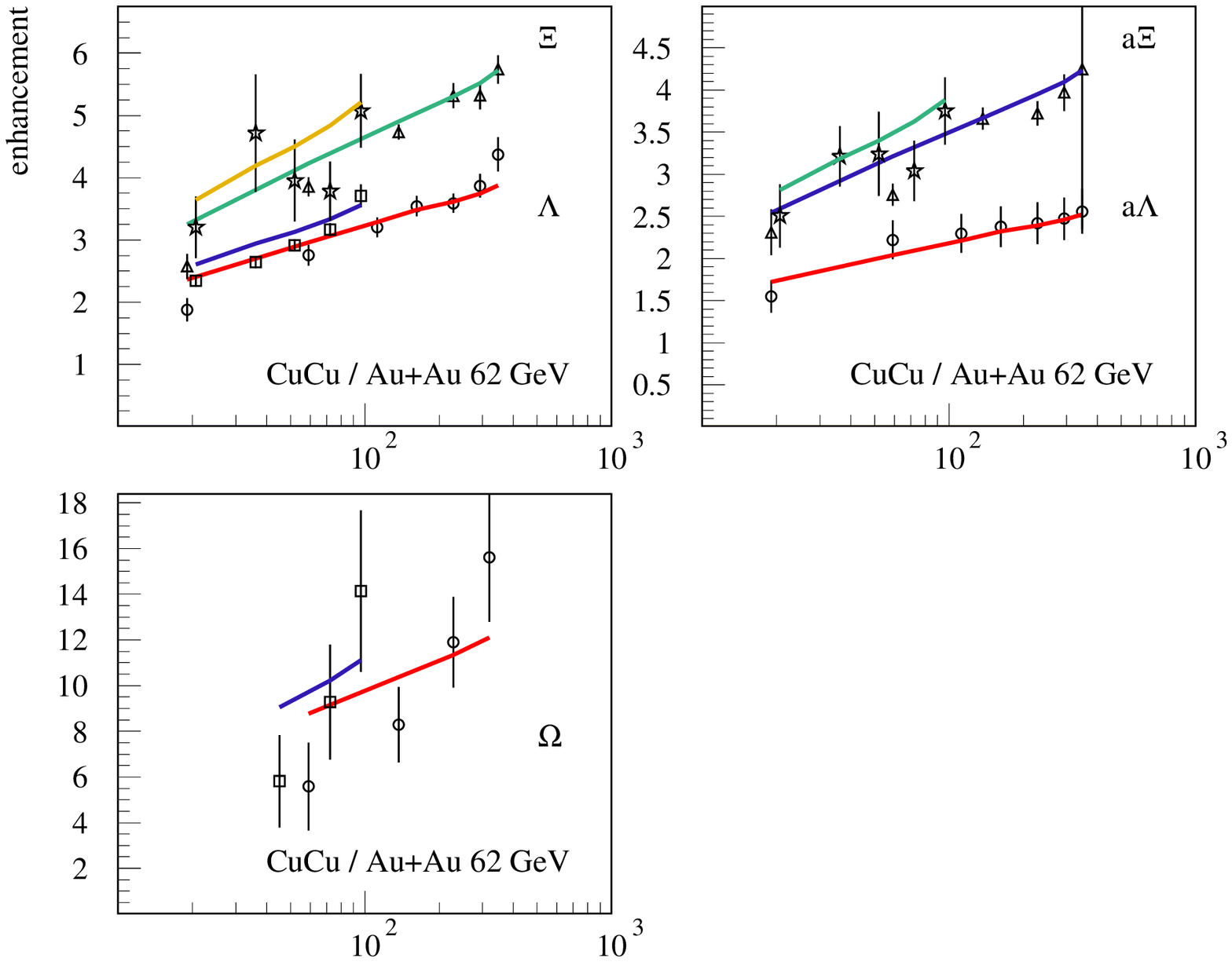}
\vspace*{-1.1cm}\par
\hspace*{-0.5cm}
\includegraphics[scale=.42,angle=-90]{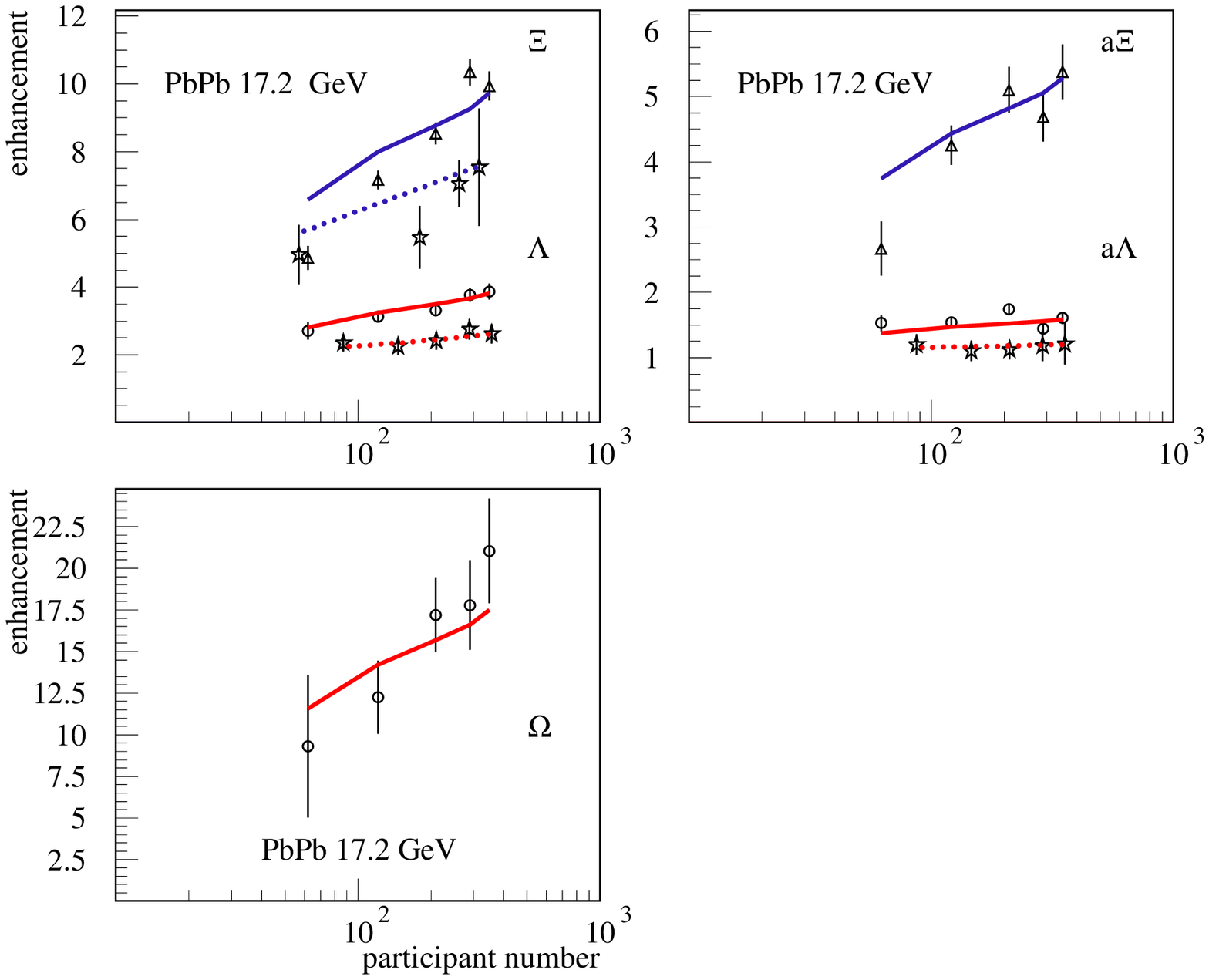}
\end{center}
\caption{ Enhancement of nuclear strange particle production for
different energies/systems. We compare toy model predictions (lines
(dashed lines for the NA49 results)) and data (points).} \label{ex}
\end{figure}
The results of our calculation (eq.\ref{eq1}) are displayed in fig.
1 for reactions at $\sqrt{s}=200 \ {\rm AGeV}, \sqrt{s}=62 \ {\rm
AGeV}$, and $\sqrt{s}=17.2 \ {\rm AGeV}$, respectively. We plot the
so-called nuclear enhancement, i.e. the ratio of the multiplicity
per participant to half the multiplicity in proton-proton
scattering, as a function of $N_{wound}$ \cite{tak} for different
hadrons and different systems. It is evident that the simple toy
model, i.e. eq. \ref{eq1}, describes  the
experimentally observed centrality dependence for all three energies, for the strange
baryons as well as for the $\phi$. Of course the assumptions of an
abrupt transition between string fragmentation and fireball
formation is rather crude but the large experimental error-bars make
it impossible to refine this model. At SPS energies data sets from 2
experiments, NA49 and NA57, are available. They do not agree. It is
therefore useful to display calculations for both data sets. Our
calculation reproduces the NA49 data but is for peripheral reactions
less steep than the NA57 data set.
\begin{figure}[ht]
\begin{center}
\hspace*{-0.5cm}
\includegraphics[scale=0.65]{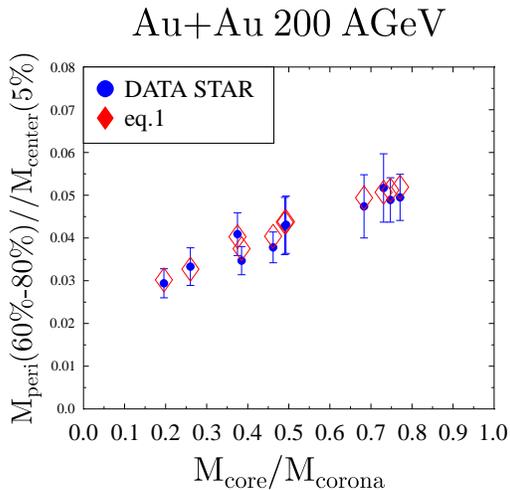}
\end{center}
\caption{Ratio of the peripheral and central multiplicity of all
particles observed in 200 AGeV Au+Au reactions as a function of the
ratio of the corona and the core multiplicity.} \label{ex1}
\end{figure}
\begin{figure}[ht]
\begin{center}
\hspace*{-0.5cm}
\includegraphics[scale=0.45,angle=-90]{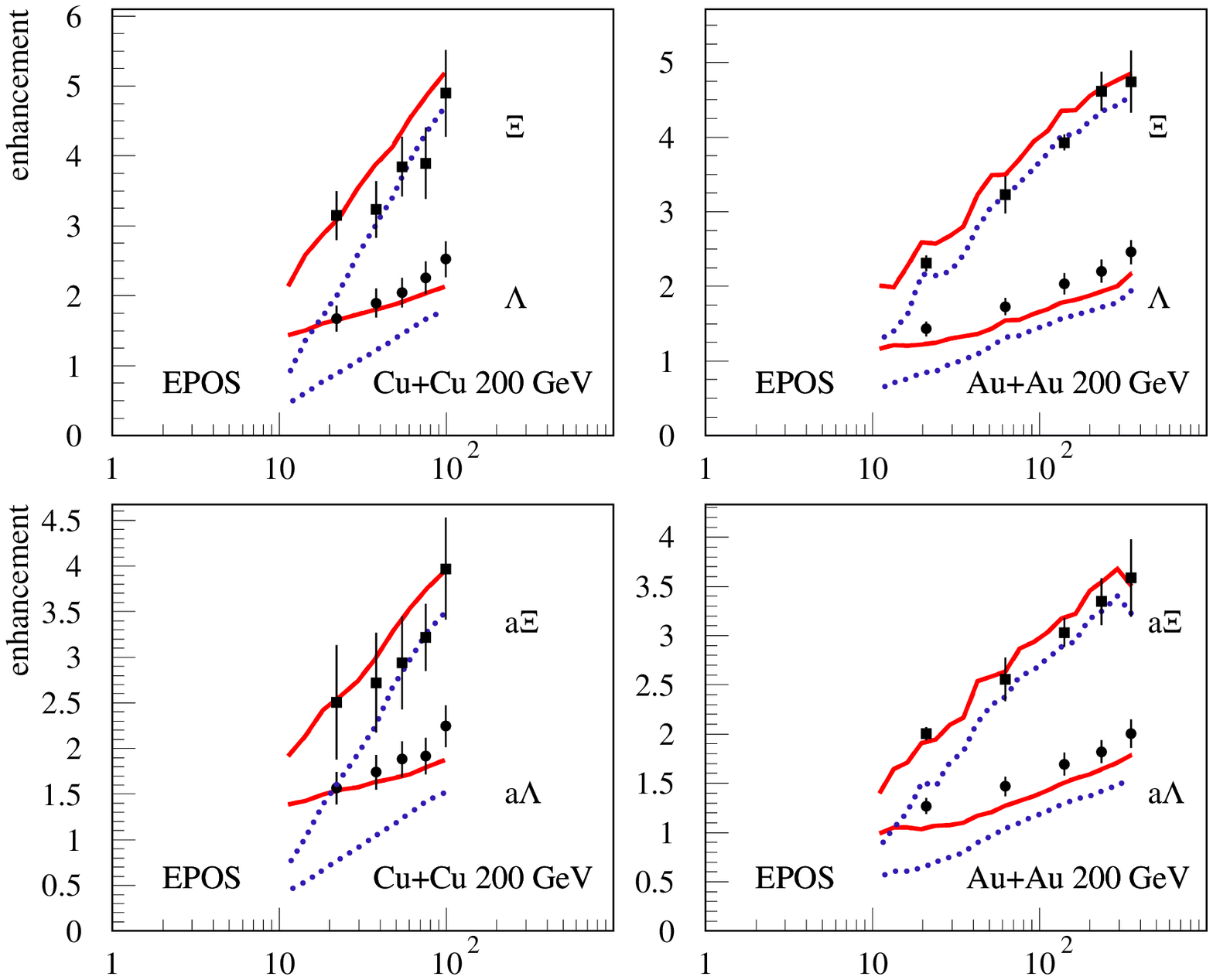}
\vspace*{-1.1cm}\par \hspace*{-0.5cm}
\includegraphics[scale=0.45,angle=-90]{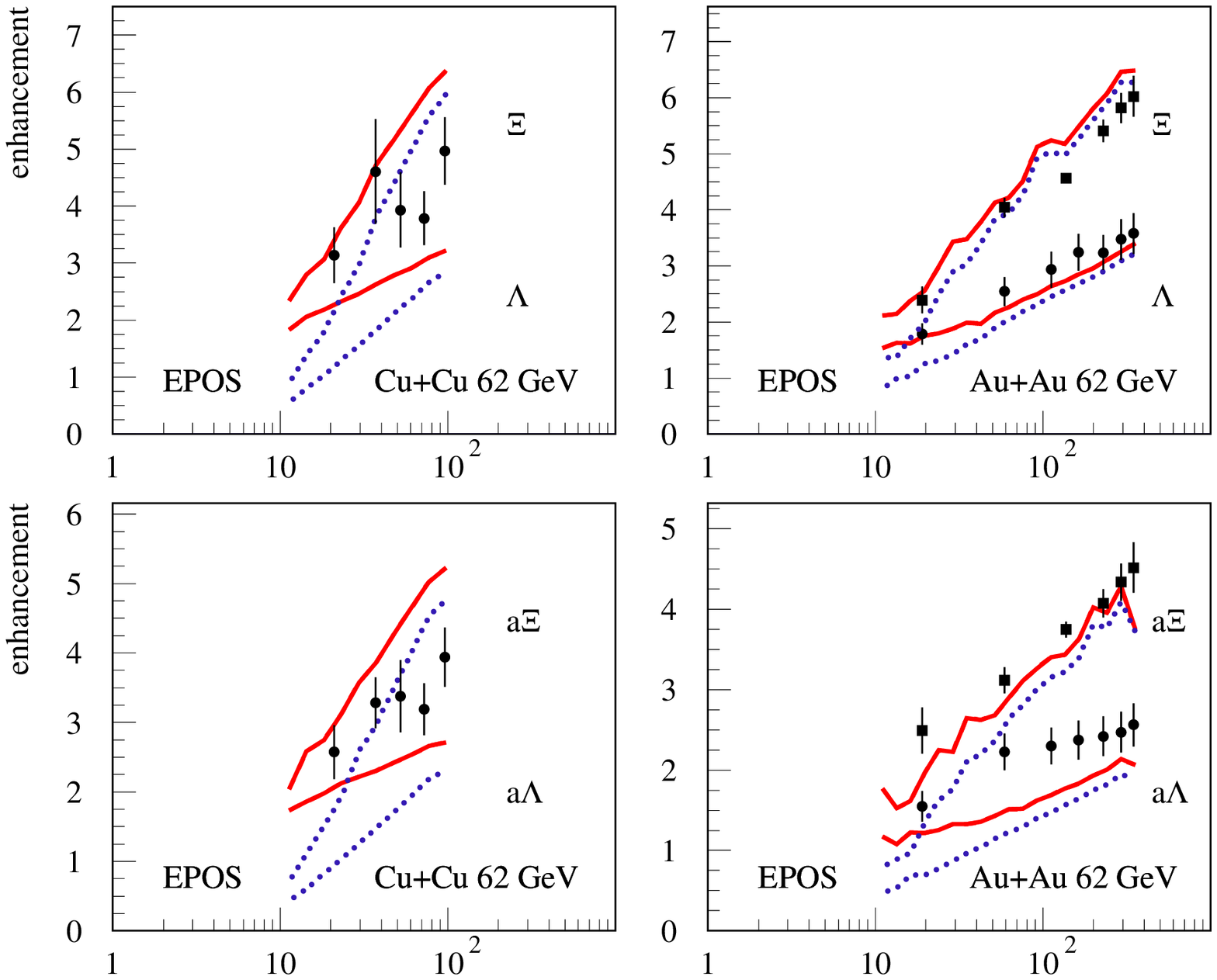}
\vspace*{-1.1cm}\par \hspace*{-0.5cm}
\includegraphics[scale=0.45,angle=-90]{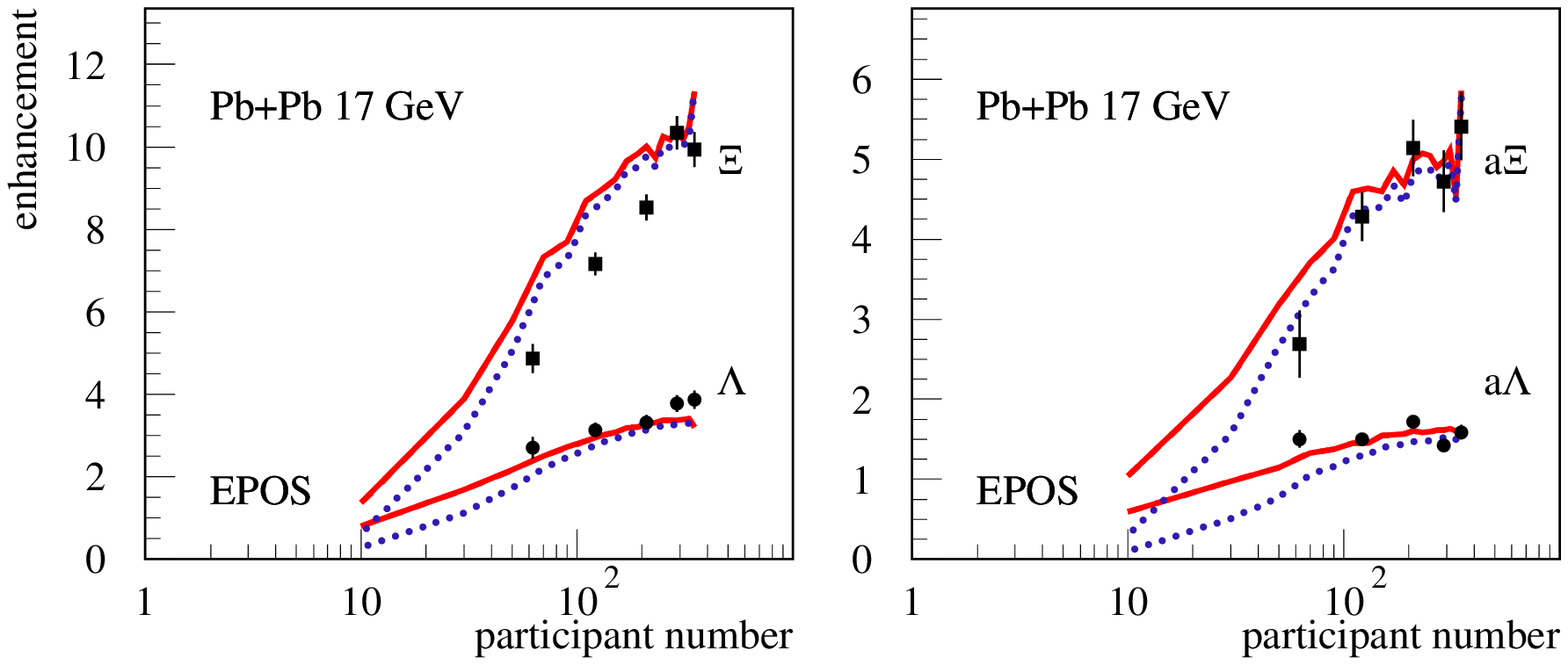}
\end{center}
\caption{Enhancement of nuclear strange baryon production for
different energies/systems. We show the full EPOS 1.9 results  as well
as the core contribution (dotted). Data are displayed as points.}
\end{figure}

It is interesting to display the core corona effect in a different
way. In fig \ref{ex1} we display the multiplicity per participating
nucleon of all observed hadrons in peripheral reactions normalized
to the same multiplicity in the most central collisions as a
function of $ M^i_{corona }/M^i_{core}$. We see that all,
non-strange as well as strange, particles fall on the same curve.
More precisely, both theory and experiment show exactly the same
linear increase.
 This means that the
enhancement of the multiplicity in central heavy ion collisions as
compared to peripheral reactions is for all particles strongly
correlated with the multiplicity difference in pp collisions as
compared to what is expected in a purely thermal environment.

 Let us now turn to a somewhat more sophisticated
implementation of the core-corona picture in the framework of the
EPOS approach \cite{klaus}. EPOS is a parton model, so in case of a
nucleus-nucleus collision, there are many binary interactions, each
one represented by a parton ladder\cite{klaus2,klaus3}. Such a
ladder may be considered as a longitudinal color field, conveniently
treated as a relativistic string. In case of proton-proton
scattering, the strings decay via the production of quark-antiquark
pairs, creating in this way string fragments -- which are usually
identified with hadrons. When it comes to heavy ion collisions, the
procedure is modified: one considers the situation at an early
proper time $\tau_{0}$, long before the hadrons are formed: one
distinguishes between string segments in dense areas (more than some
critical density $\rho_{0}$ segments per unit volume), from those in
low density areas. The high density areas are referred to as core,
the low density areas as corona. Compared to our toy model, here we
perform a real three-dimensional core/corona separation: the core
contribution will be more important at central rapidities as
compared to  projectile and target rapidities.

Whereas corona particle production is as in proton-proton
scattering, the core part is assumed to be collectively expanding:
Bjorken-like in longitudinal direction with in addition some
transverse expansion. We assume particles to freeze out at some
given energy density $\varepsilon_{\mathrm{FO}}$, having acquired at
that moment a collective radial flow. The latter one is
characterized by a linear radial rapidity profile from inside to
outside with maximal radial rapidity $y_{\mathrm{rad}}$ (model
parameter). Hadronization then occurs according to covariant
microcanonical phase space. So the core definition and its
hadronization are parameterized in terms of few global parameters,
for details see  \cite{klaus}.

In fig. 3, we plot the nuclear enhancement obtained from EPOS 1.9. We observe, as
already in the simple toy model, a steeper increase of multiplicities with
centrality in case of CuCu compared to AuAu, due to a different $N_{\rm part}$
dependence of the relative core/corona weight in the two systems.

In conclusion, we have shown that the centrality dependence of
strange hadron production can be well explained as a purely
geometrical effect: the total volume is composed of a low density
part -- the corona, and a high density part -- the core. Particle
production in the corona is  essentially string breaking as in pp
collisions, whereas the core represents a fireball from which
hadrons are produced according to phase space as in statistical
models.
 This approach explains in a natural way the striking difference in the
 centrality dependence in CuCu compared to AuAu, being simply due to a
 different relative core/corona weight at a given $N_{\rm part}$ for the two
 systems.

 The $\phi$ mesons fit perfectly into this picture, the difference with respect
 to other hadrons like $\Omega$ or $\Xi$ is quantitative. The ratio of core to
 corona multiplicity is considerably bigger for the latter ones, therefore the
 enhancement  increases faster with $N_{\rm part}$ than in case of  $\phi$.
 As a consequence, the $\phi$ enhancement for CuCu is close to the AuAu result.

  We have demonstrated that a very simple core/corona toy model already
 explains the main features seen in the data. More sophisticated EPOS
 simulations confirm these finding \cite{kk}.

 It is a consequence of this
 observation that the production of strange baryons -- as that of all other hadrons --
 follows very simple rules: after subtracting the corona part (whose importance
 depends on centrality / system size), all particles are created from some bulk
 matter whose properties are independent of the size of the system.
 It is actually remarkable that the core part is strictly
 determined by geometry and becomes therefore very small for the very peripheral
 reactions. This continuity implies that the expected new state of matter
 would exists also in very small systems.
 Another important consequence of this observation is that the hadron yield
 of hadrons containing u, d, and s quarks only, is sensitive to the freeze out
 of the matter. If one wants to have information on earlier stages other
 probes have to be used.

{\bf Acknowledgement:} We would like to thank Dr. Timmins for
sending us the Glauber model values of the fraction of corona
particles and the Star collaboration, especially Drs. Bedanga,
Blume, Takahashi and Munhoz, for communicating to us the preliminary
multiplicity values of the Pb+Pb and Cu+Cu run.


\begin{thebibliography}{99}
\bibitem{kk} In this framework, adding a final state
 hadronic cascade, we find that the corona part actually survives the space
 time evolution
\end{thebibliography}
\end{document}